# Systematic Transaction Level Modeling of Embedded Systems with SystemC

Wolfgang Klingauf
Technical University of Braunschweig, Abt. E.I.S.
Muehlenpfordtstr. 23, 38106 Braunschweig, Germany
klingauf@eis.cs.tu-bs.de

**Abstract**

*This paper gives an overview of a transaction level modeling (TLM) design flow for straightforward embedded system design with SystemC. The goal is to systematically develop both application-specific HW and SW components of an embedded system using the TLM approach, thus allowing for fast communication architecture exploration, rapid prototyping and early embedded SW development. To this end, we specify the lightweight transaction-based communication protocol SHIP and present a methodology for automatic mapping of the communication part of a system to a given architecture, including HW/SW interfaces.*

## 1. Introduction

Recently, the transaction level modeling (TLM) paradigm has been widely propagated for System-on-Chip (SoC) design. By orthogonalizing system functionality and system communication, very high simulation speeds become feasible enabling fast communication architecture exploration [4].

In contrast to SoC modeling, the design of embedded systems typically incorporates the assembly of standard HW and SW components with user-designed HW (reconfigurable logic or ASIC) and SW. As system complexity continuously rises, the proper connection of user HW and SW to the systems communication architecture becomes more and more a focus of design. As a result, the development of embedded software (eSW) that is closely related to the HW will have to wait for the RTL model to be completed.

To fill this gap, we present a TLM approach for embedded system design with SystemC that considerably relieves the designer of the task of implementing platform-specific communication protocols. Moreover, fast communication architecture exploration, early eSW development, and rapid prototype generation is supported.

The proposed design flow is displayed in figure 1. It specifies three TLM models, namely component-assembly model (defined in [2]), cycle count accurate at the boundaries model (CCATB, introduced in [4]), and communication architecture model (adapted from cycle-accurate computation model in [2]).

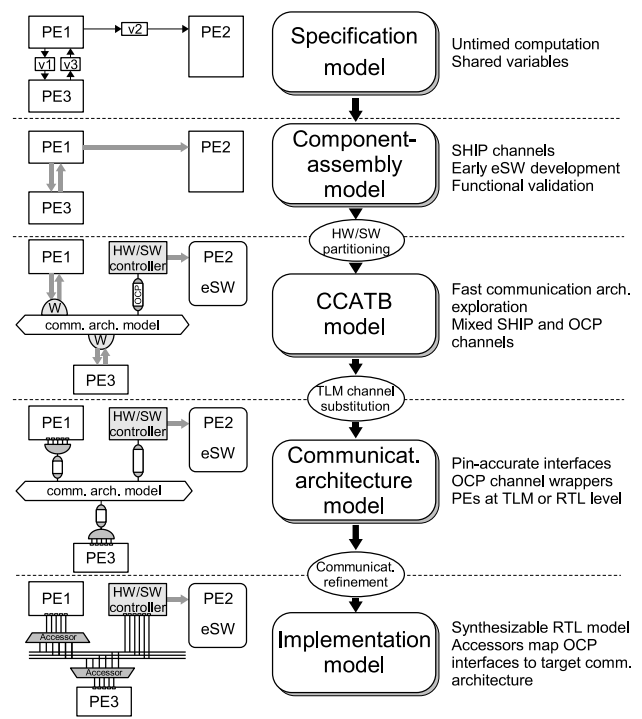

**Figure 1. Design flow**

To encourage highly systematic development, each model is clearly associated with a predetermined communication protocol. For communication modeling in the upper two TLM models, we introduce the SystemC High-level Interface Protocol (SHIP). Below the CCATB model, the widely supported and openly-licensed Open Core Protocol (OCP) [1] is used. By specifying a generic SHIP-based HW/SW interface, fully transaction-based HW/SW communication in the final system is rendered possible.

## 2. SHIP

SHIP is a lightweight communication protocol for transaction-based modeling of directed point-to-point connections between two communication entities such as processing elements (PE) or communication architecture accessors. SHIP was designed to allow for top-down communication modeling on a level of abstraction that is completely independent of the HW/SW partitioning.

The SHIP channel is a message passing channel which transfers any C++ object that implements the `ship_serializable_if` interface. This interface defines the `serialize` and the `deserialize` function. The channel calls these functions to transform communication objects into serial data streams and vice versa.

The SHIP channel offers four blocking interface method calls: `send`, `recv`, `request`, and `reply`. While PEs that exclusively use the `send` and `request` functions implicitly represent a communication *master*, `recv` and `reply` are *slave* methods. When consequently applied, this allows for automatic master/slave detection.

## 3. Communication architecture exploration and prototyping

In the following, we briefly describe our concept of fast communication architecture exploration and prototyping. It is based on communication architecture models (CAM) and accessors.

**Communication architecture model.** With this term, we denote a simulation model of a communication architecture such as a bus or a network. CAMs are CCATB models with a cycle-accurate notion of time when viewed at transaction boundaries. Internally, only timed method calls are used which reflect the simulated bus or network protocol. Given a library of CAMs (e.g. of the CoreConnect architecture), fast yet timing-accurate communication architecture exploration is feasible.

It should be noted that the CAMs can be utilized in all abstraction models below the component-assembly model. They are connected to the systems PEs using OCP TLM interfaces. By the use of wrappers, virtually any PE can be connected to the CAM, independent of its communication interface. In our approach, wrappers for high-level SHIP and pin-accurate OCP interfaces are provided.

**Communication architecture accessors.** They are intended for the automatic generation of a synthesizable prototype of the hardware part. Their use implies that the designer has refined all PEs to the RTL level and has implemented a pin-level OCP interface. Then, to connect a PE to a selected target communication architecture (e.g. CoreConnect), the appropriate accessor is attached to the PE. Since accessors are implemented as RTL, they are fully synthesizable.

## 4. SW synthesis and HW/SW communication

The ultimate goal of the proposed design methodology is to use SystemC as a unifying system specification language and, after HW/SW partitioning, to generate eSW automatically from the SystemC code. Moreover, HW/SW communication should be established without requiring any changes to the source code. To this end, we adopt the methodology presented in [3]. Embedded SW can be systematically generated from SystemC code by simply substituting some SystemC library elements for behaviourally equivalent procedures based on RTOS functions.

However, this methodology does not include HW/SW interface generation. Hence, we define two constraints: First, eSW generation takes place in a transaction-level model of the system, namely the component-assembly model. Second, the PEs that are to become eSW exclusively must use SHIP channels for communication with other PEs of the system. PEs that fulfill these requirements can immediately be synthesized to eSW entities.

To enable fully transaction-based HW/SW communication, we specify a generic HW/SW interface supporting SHIP-based communication. This interface virtually realizes a SHIP channel with one end in the HW partition and one end in the SW partition of the system. Its implementation splits up into a HW and a SW adapter.

The HW adapter essentially features a pin-level OCP interface that allows for connecting to the systems communication architecture. Data exchange with the SW adapter is implemented by shared memory and sideband signals.

The SW part of the HW/SW interface consists of a device driver and a small communication library. While handshaking and memory-mapping is accomplished by the device driver, the communication library implements the SHIP channel interface method calls.

## 5. Conclusion

Currently, the design flow is being implemented for the IBM CoreConnect architecture and embedded Linux OS. A first case example will be demonstrated at the interactive presentation for which this paper is a base. Future research will include practical and theoretical analysis of limits and application areas of the methodology.